\documentclass[floatfix,aps,amsmath,amssymb,reprint,twocolumn,showpacs,showkeys,superscriptaddress]{revtex4-2}
\UseRawInputEncoding

\usepackage{amsmath,amsfonts,amsthm,amssymb,array,bm,braket,dcolumn,enumerate,flafter,float,graphics,latexsym,makeidx,multirow,natbib,verbatim,xfrac}
\usepackage[english]{babel}
\usepackage[mathscr]{euscript}
\usepackage[utf8]{inputenc}
\usepackage[usenames]{color}
\usepackage[pdftex,colorlinks=true,linkcolor=blue,urlcolor=blue,citecolor=blue]{hyperref}
\usepackage{graphicx}
\begin{document}
\title{Frequency as a Clock: Synchronization and Intrinsic Recovery in Graphene Transistor Dynamics}

\author{Victor Lopez-Richard}
\email{vlopez@df.ufscar.br}
\affiliation{Departamento de Fisica, Universidade Federal de Sao Carlos, 13565-905, Sao Carlos, SP, Brazil}

\author{Igor Ricardo Filgueira e Silva}
\affiliation{Departamento de Fisica, Universidade Federal de Sao Carlos, 13565-905, Sao Carlos, SP, Brazil}

\author{Gabriel L. Rodrigues}
\affiliation{Brazilian Nanotechnology National Laboratory (LNNano), Brazilian Center for Research in Energy and Materials (CNPEM), 13083-200, Campinas, São Paulo, Brazil}
\affiliation{“Gleb Wataghin” Institute of Physics, State University of Campinas, 13083-970 Campinas, São Paulo, Brazil}

\author{Rafael Furlan de Oliveira}
\affiliation{Brazilian Nanotechnology National Laboratory (LNNano), Brazilian Center for Research in Energy and Materials (CNPEM), 13083-200, Campinas, São Paulo, Brazil}

\author{Kenji Watanabe}
\affiliation{Research Center for Electronic and Optical Materials, National Institute for Materials Science, 1-1 Namiki, Tsukuba 305-0044, Japan}

\author{Takashi Taniguchi}
\affiliation{Research Center for Materials Nanoarchitectonics, National Institute for Materials Science, 1-1 Namiki, Tsukuba 305-0044, Japan}

\author{Alisson R. Cadore}
\email{alisson.cadore@lnnano.cnpem.br}
\affiliation{Brazilian Nanotechnology National Laboratory (LNNano), Brazilian Center for Research in Energy and Materials (CNPEM), 13083-200, Campinas, São Paulo, Brazil}
\affiliation{Programa de Pós-Graduação em Física, Instituto de Física, Universidade Federal de Mato Grosso, Cuiabá, Brazil}

\begin{abstract}
Hysteresis and memory effects in graphene field-effect transistors (GFETs) offer unique opportunities for neuromorphic computing, sensing, and memory applications, yet their physical origins remain debated due to competing volatile and nonvolatile interpretations. Here, we present a unified dynamic model that captures the essential physics of the GFET response under periodic gate modulation, accounting for both intrinsic relaxation processes and externally driven charge transfer. By modeling non-equilibrium carrier dynamics as a competition between injection and reabsorption rates, we uncover two distinct regimes: one governed by intrinsic, frequency-independent relaxation and another exhibiting frequency-locked behavior where the response is tied to the external drive. This distinction resolves apparent nonvolatile effects and explains loop invariance in floating-gate structures via displacement current–driven charge injection. Our framework predicts the evolution of the hysteresis loop shape, amplitude, and direction across a wide range of driving conditions, offering a versatile tool for interpreting experimental results and guiding the design of next-generation graphene-based electronic systems.
\end{abstract}

\maketitle

\section{Introduction}

Graphene field effect transistors (GFETs) are highly promising devices that take advantage of the exceptional electronic properties of graphene, such as ultrahigh carrier mobility, atomic thickness, and mechanical flexibility, making them ideal candidates for high frequency communication~\cite{Viti2021,Behera2024}, advanced computing~\cite{Soliman2025}, and highly sensitive sensing applications~\cite{Cadore2016,Lou2025}. Despite challenges such as the absence of an intrinsic bandgap, GFETs continue to show significant potential for next-generation electronics~\cite{Garcia2025}.

A particularly intriguing aspect of GFET operation lies in the emergence of conductance memory effects, often manifested as hysteretic current–voltage loops. These effects enable functionalities such as neuromorphic computing~\cite{Yao2019,Soliman2025} and memory storage~\cite{wang2019new,Selmi2024}, and are typically attributed to competing mechanisms, including surface dipoles, interfacial traps, and ferroelectric polarization~\cite{kurchak2017hysteretic}. Experimental studies~\cite{wang2010hysteresis,cadore2016thermally} have shown that such hysteresis could strongly depend on the gate voltage sweep rate, amplitude, and dielectric environment, suggesting a dynamic and rate-sensitive nature. 

Previous models of floating gate (FG) transistors have primarily relied on Fowler–Nordheim (FN) tunneling as the dominant charging mechanism~\cite{fu2025non}, without accounting for capacitive coupling. However, such approaches lead to contradictions with our experiments: (i) when instantaneous currents depend only on the applied voltage, their time integration to obtain the accumulated charge introduces an unavoidable dependence on the voltage sweep rate, in contrast to our measurements, and (ii) they result in residual charge accumulation after each cycle, preventing the stable consecutive operation we observe~\cite{Ziegler2012}. Other models, such as those in Ref.~\citenum{Winters2015}, introduce hysteresis in the Fermi energy by invoking unspecified resonant processes at the graphene/dielectric interface, but fail to clarify the microscopic origin of these effects. In contrast, our results reveal a direct and unambiguous correlation between drain and gate currents, which these models overlook. More recent works on van der Waals heterostructures~\cite{Liu2021,Wang2024,Shi2020,Soliman2025} reduce the operation mechanism to intuitive, threshold-based charge transfer diagrams, without developing a self-consistent analytical or numerical treatment. Crucially, none of these approaches account for the polarity independence of our response or the observed insensitivity of gate charging to sweep rate. These discrepancies highlight the need for a refined framework: one capable of reconciling volatile and non-volatile behaviors under a unified, dynamic picture. Our work takes a step in this direction, while also pointing toward the necessity of deeper microscopic investigations beyond the current state of the art. 

Additionally, dynamic models based on differential equations are also essential to capture the transient and frequency-dependent behavior of nonlinear electronic devices, going beyond static or linear-impedance descriptions~\cite{LopezRichard2024b}. By incorporating the explicit time dependence of charge, current, and voltage, they provide a more reliable framework for predicting device–circuit interactions and overall circuit performance~\cite{LopezRichard2024a,LopezRichard2024}.

Motivated by these challenges, this work proposes a unified phenomenological framework that elucidates the key physical mechanisms driving GFET dynamics, delivering predictive and interpretative insight across a broad range of experimental conditions, while deliberately avoiding the ambition of a full microscopic treatment or strict quantitative matching with experiments. By linking loop characteristics, such as shape, direction, and size, to the amplitude and period of the applied gate voltage, the model distinguishes between two fundamental regimes: one governed by intrinsic, frequency-independent resistive relaxation, and another dominated by capacitive charge exchange  locked to the external drive. This distinction provides insight into the physical origins of volatile versus non-volatile behavior in graphene-based devices.

\section{Results and Discussion}

\begin{figure}[!htb]
    \centering
\includegraphics[width=1\linewidth]{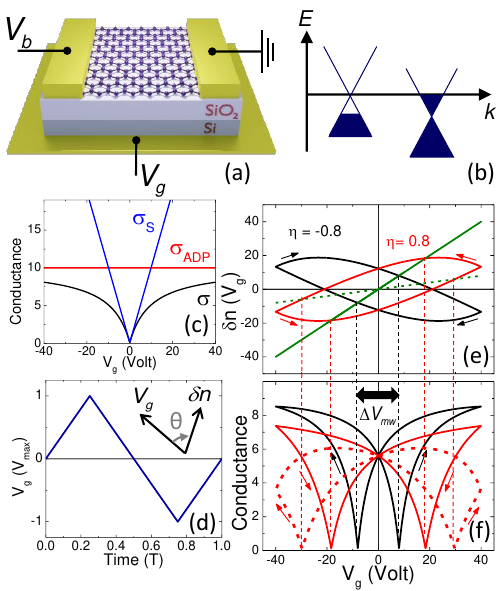}
\caption{(a) Schematic of a graphene field-effect transistor fabricated on a SiO$_2$/Si substrate.  
(b) Illustration of graphene’s electronic structure modulation under gating, either via Fermi level shift or band bending.  
(c) Effective conductance $\sigma$ as a combination of sheet conductance $\sigma_S$ and acoustic deformation potential scattering contribution $\sigma_{ADP}$.  
(d) Gate voltage sweep applied to the device and their phasor representation, shown relative to the resulting carrier density fluctuation.  
(e) Charge fluctuation versus gate voltage loops illustrating distinct charge transfer mechanisms. Solid and dashed green lines correspond to different values of $c_g V_g$.  
(f) Corresponding conductance versus gate voltage loops derived from the charge modulation behavior in (e).
}
    \label{fig:01}
\end{figure}

In order to characterize charge transport through the graphene layer (Fig.~\ref{fig:01} (a)), we consider the sheet conductance given by $\sigma_S = \mu n(V_g)$, where charge carriers experience scattering dominated by long-range Coulomb interactions with impurities~\cite{Stauber2007,Silvestre2013} leading to mobility $\mu$. We may also asume a linear dependence of the charge density on gate voltage, expressed as~\cite{Chen2008}
\begin{equation}
    n(V_g) = c_g |V_g - V_{g,\text{min}}|,
\end{equation}
where $V_{g,\text{min}}$ denotes the gate voltage corresponding to the charge neutrality point and $c_g$, an effective capacitance. As illustrated in Fig.~\ref{fig:01} (b), this neutrality point can shift either due to band filling or depletion (resulting in a shift of the Fermi level), or due to electrostatic band bending. Both mechanisms can be attributed to nonequilibrium charge fluctuations, $\delta n$, occurring either within the graphene layer or in surrounding regions. Accordingly, the shift in the neutrality point can be described by the relation $V_{g,\text{min}} = \delta n(V_g)/c_g$.
Additionally, we consider that the total conductance is limited by the contribution from interactions with acoustic phonons via the deformation potential, denoted as $\sigma_{ADP}$. This contribution is independent of band filling~\cite{Stauber2007,Hwang2008} and, therefore, unaffected by gating, as illustrated in Fig.~\ref{fig:01} (c).

According to Matthiessen’s rule, the total resistance can be obtained by summing the individual contributions from different scattering mechanisms
\begin{equation}
    \frac{1}{\sigma} = \frac{1}{\sigma_S} + \frac{1}{\sigma_{ADP}} + \rho_c,
\end{equation}
where $\rho_c$ represents an unavoidable parasitic contact resistance. This leads to the following expression for the total conductance
\begin{equation}
    \sigma = \sigma_{\text{res}} + \frac{\mu n(V_g)}{1 + \tilde{\rho} \mu n(V_g)},
    \label{cond}
\end{equation}
with the charge density defined as
\begin{equation}
    n(V_g) = \left| c_g V_g - \delta n(V_g) \right|,
    \label{dens}
\end{equation}
and the gate-independent resistive factor given by
\begin{equation}
    \tilde{\rho} = \frac{1}{\sigma_{ADP}} + \rho_c.
\end{equation}
The term $\sigma_{\text{res}}$ in Eq.~\ref{cond} accounts for the residual conductance arising from leakage channels, which serves to regularize the otherwise unphysical divergence of resistance as $V_g \rightarrow V_{g,\text{min}}$. The resulting total conductance is illustrated in Fig.~\ref{fig:01} (c) by the black curve, exhibiting the characteristic V-shaped dependence on gate voltage. 
This expression can be further refined by treating electrons ($V_g > V_{g,\text{min}}$) and holes ($V_g < V_{g,\text{min}}$) separately, allowing for distinct values of mobility and contact resistance for each carrier type. However, for simplicity, all calculations presented here assume a symmetric case.


Subsequently, upon applying a time-dependent gate voltage $V_g(t)$, the device exhibits a dynamic response governed by the nonequilibrium charge (or electric polarization) fluctuations $\delta n(t)$. A common experimental approach for characterizing this behavior is the application of triangular voltage sweeps, as illustrated in Fig.~\ref{fig:01} (d), which are typical gate voltage sweeps in transistor measurements. Alternatively, sinusoidal sweeps can also be employed, allowing both the gate voltage and the resulting charge fluctuations to be conveniently represented as phasors, as shown in the inset of Fig.~\ref{fig:01} (d). 

In this work, we examine both types of excitation; however, for clarity and closer alignment with experimental conditions, our graphical representations will focus primarily on triangular waveforms. This choice facilitates a more direct correlation between theoretical predictions and observed device behavior.
\begin{figure*}[!htb]
    \centering
    \includegraphics[width=1\linewidth]{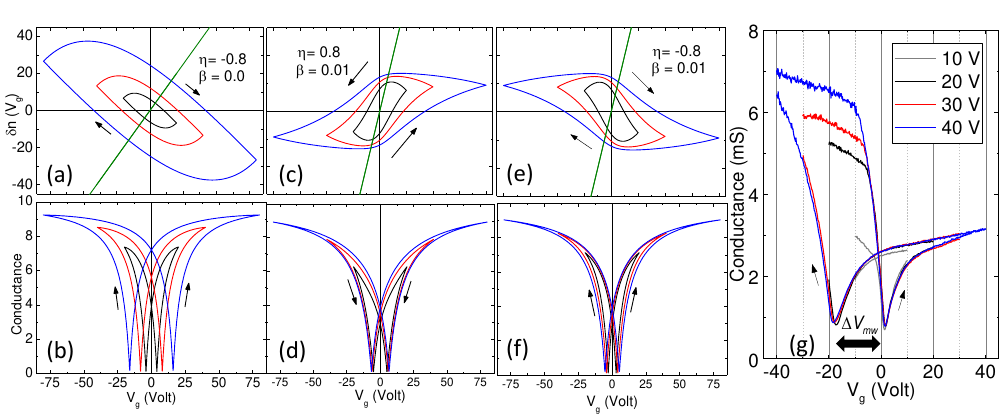}
    \caption{(a) Theoretically computed charge fluctuation versus gate voltage loops as a function of pulse amplitude, in the absence of saturation effects. The green solid line indicates the ideal linear relation $c_g V_g$.  
(b) Corresponding conductance versus gate voltage loops derived from the carrier density profiles in (a).  
(c) Charge fluctuation versus gate voltage loops exhibiting saturation or exhaustion behavior due to limited charge transfer.  
(d) Conductance versus gate voltage loops corresponding to the saturated profiles in (c).  
(e,f) Results for a contrasting charge transfer mechanism, showing both carrier density (e) and the associated conductance response (f).  (g) Experimentally measured conductance versus gate voltage loops under varying bottom-gate voltage amplitudes, displaying minimal sensitivity to amplitude changes. Panel (g) corresponds to a GFET without top gates.
}
    \label{fig:02}
\end{figure*}
The simplest approach to characterize the dynamics of nonequilibrium charge fluctuations under a time-varying gate voltage $V_g(t)$ is to assume that the evolution of $\delta n(t)$ is governed by two competing transfer rates following the equation
\begin{equation}
    \frac{d \delta n}{dt} = g_{\text{out}}(V_g) + g_{\text{in}}(V_g),
\end{equation}
where $g_{\text{out}}(V_g)$ represents the charge recombination or reabsorption rate, and $g_{\text{in}}(V_g)$ denotes the injection rate, with the physical nature of each process inferred from its sign. Under equilibrium conditions, these rates are constrained by detailed balance, such that $g_{\text{out}}(V_g) = - g_{\text{in}}(V_g)$. It is important to emphasize that these charge transfer processes need not occur directly within the graphene layer. A purely capacitive coupling to localized states or confinement sites, such as those at nearby interfaces or in the dielectric environment, is sufficient to induce a fluctuation in the relative position of the Fermi level with respect to the band center as represented in Fig.~\ref{fig:01} (b). These fluctuations, in turn, modulate the carrier density (Eq.~\ref{dens}) and affect the overall device response (Eq.~\ref{cond}).


\begin{figure*}[!htb]
    \centering
    \includegraphics[width=1\linewidth]{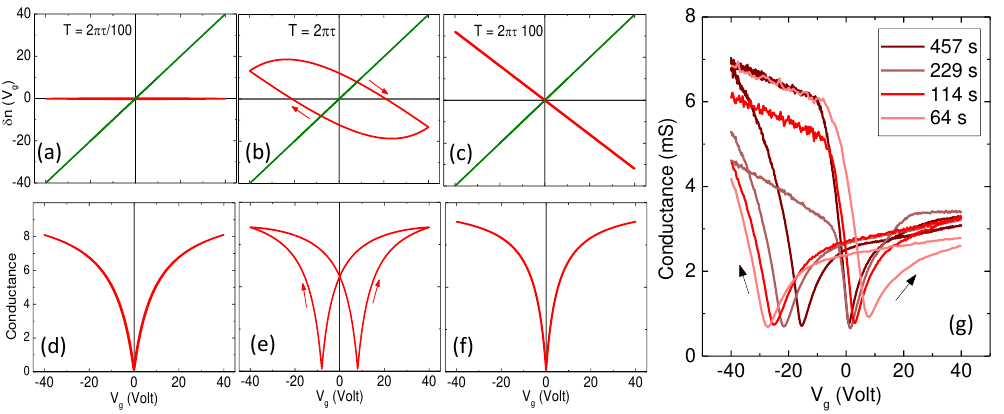}
    \caption{Theoretically computed charge fluctuation versus gate voltage loops as a function of pulse period.  
The green solid line represents the linear relation $c_g V_g$.  
(a) $T=2 \pi \tau/100$,  
(b) $T=2 \pi \tau$, and (c) $T=2 \pi \tau\cdot 100$.  
(d)–(f) Corresponding conductance versus gate voltage loops derived from the carrier density profiles in (a), (b), and (c), respectively.  
(g) Experimentally measured conductance versus gate voltage loops under varying sweep periods, demonstrating a pronounced frequency dependence of the hysteresis behavior. Panel (g) corresponds to a GFET without top gates.}
    \label{fig:03}
\end{figure*}
If the dynamic response of charge carriers to external perturbations is volatile and predominantly governed by an intrinsic relaxation time $\tau$, then, for sufficiently small deviations from equilibrium, the recombination (or reabsorption) rate can be approximated as  
\begin{equation}
    g_{\text{out}} = -\frac{\delta n}{\tau}.
    \label{gout1}
\end{equation}
Meanwhile, within a certain range of applied gate voltages, the charge transfer due to carrier generation or trapping can be reasonably assumed to vary linearly with the gate voltage, particularly in scenarios where polarity-dependent behavior is expected
\begin{equation}
    g_{\text{in}} = \eta V_g.
    \label{gin1}
\end{equation}
Here, the sign of the proportionality constant $\eta$ determines the dominant transfer mechanism. Specifically, $\eta > 0$ and $\eta < 0$ correspond to opposite directions of the resulting hysteresis loops.

Two calculated $\delta n(V_g)$ loops, circulating in opposite directions due to the sign of $\eta$, are shown in Fig.~\ref{fig:01} (e), with the corresponding conductance loops plotted in Fig.~\ref{fig:01} (f). According to Eq.~\ref{dens}, the position of the V-shaped conductance minimum is determined by the intersection of each $\delta n(V_g)$ loop with the green dashed line defined by $c_g V_g$, which is also indicated in Fig.~\ref{fig:01} (e). The V-shaped minimum exhibits a clear shift between the forward and reverse voltage sweeps in all cases. This shift in gate voltage, denoted as $\Delta V_{mw}$ and highlighted in Fig.~\ref{fig:01} (f), is referred to as the memory window and will serve as a measure of the strength of the memory effect. By reversing the sign of $\eta$, the positions of the V-shape minima change. However, the direction of the conductance loop alone is not a definitive indicator of the underlying charge transfer mechanism. A more nuanced analysis involves comparing conductance responses derived from the same $\delta n(V_g)$ loop but under different values of $c_g V_g$, represented by the green solid and green dashed lines in Fig.~\ref{fig:01} (e). These result in the respective red solid and red dashed conductance curves shown in Fig.~\ref{fig:01} (f), which exhibit clearly opposite loop directions, despite originating from the same charge fluctuation dynamics. 

Under these conditions, the system’s response remains sensitive to variations in the gate voltage amplitude, as illustrated in Fig.~\ref{fig:02}(a) and (b), which display the $\delta n(V_g)$ loops and their corresponding conductance traces, respectively. However, a distinct physical effect can significantly alter this behavior: the onset of exhaustion in charge transfer processes, or saturation in cases involving electric polarization. As described in Ref.~\citenum{silva2025microscopic}, such effects can be modeled by modifying the injection term in Eq.~\ref{gin1} as $\eta \rightarrow \eta/\sqrt{1 + \beta V_g^2}$. The resulting dynamics for finite values of $\beta$ are shown in Figs.~\ref{fig:02} (c-d) and (e-f), corresponding to opposite signs of $\eta$.

As $\beta$ increases, the $\delta n(V_g)$ loops exhibit saturation, leading to a vertical narrowing or ``condensation'' of the loops across varying gate voltage maxima. This behavior results in a reduced sensitivity of the memory window in the conductance curves with respect to the applied amplitude. An experimental observation displaying similar characteristics is shown in Fig.~\ref{fig:02} (g), corresponding to a graphene-based device studied in Ref.~\citenum{cadore2016thermally}. In this work, Cadore and co-workers report on the gate hysteresis of graphene resistance in single bottom-gate graphene/hexagonal boron nitride (hBN) devices operated at temperatures above 375 K. The results clearly indicate that the memory window (hysteresis loop) exhibits negligible dependence on voltage amplitude. Additional experimental details are provided in the Methods section. Similarly, Wang et al.~\cite{wang2010hysteresis} also demonstrated the invariance of $\Delta V_{mw}$ with respect to bias range in electrolyte-gated graphene devices.

Such a volatile response is also characterized by its dependence on the voltage sweep period or, equivalently, the driving frequency. This behavior is illustrated in the $\delta n(V_g)$ loops and their corresponding conductance traces shown in Figs.~\ref{fig:03}~(a)–(c) and (d)–(f), respectively, for increasing sweep periods. A hallmark of volatile dynamics is the existence of an optimal frequency on the order of $T \sim 2\pi\tau$, around which the hysteresis loop reaches its maximum area. For frequencies much lower or higher than this characteristic timescale, the loops tend to collapse~\cite{silva2022,lopezRichard2022}.
Although, Refs.~\citenum{wang2010hysteresis} and ~\citenum{cadore2016thermally} on graphene-based devices have not reported this non-monotonic behavior (an increase followed by a reduction), it has been observed in organic-based transistors by Sun et al.~\cite{sun2015}.

This behavior can be better understood through the analytical solution of the system under a sinusoidal gate voltage of the form $V_g(t) = V_g^{\text{max}} \cos(\omega t)$. In this case, the nonequilibrium charge fluctuation follows
\begin{equation}
   \delta n(t) = \frac{\eta V_g^{\text{max}} \tau}{\sqrt{1+(\omega \tau)^2}} \cos \left(\omega t + \phi\right),
   \label{n1}
\end{equation}
where the phase shift $\phi$, depicted via phasor representation in Fig.~\ref{fig:01} (d), is given by $ \phi = -\arctan(\omega \tau)$. It is precisely this finite phase shift between the driving voltage and the induced charge fluctuation that gives rise to the hysteresis loop. Specifically, $\phi$ disappears in the low-frequency limit, and at high frequencies, $\delta n(t)$ vanishes, which strips the system of its memory effect and leads to a collapsed loop structure. The experimental demonstration of the frequency-dependent volatile behavior is shown in Fig.~\ref{fig:03} (g), which highlights the memory window's sensitivity to the applied period. This finding is consistent with observations in similar devices, as reported by Wang et al.~\cite{wang2010hysteresis}.


\begin{figure}[!htb]
    \centering
    \includegraphics[width=1\linewidth]{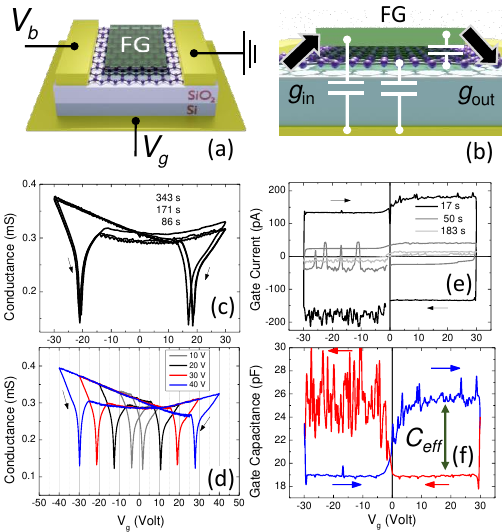}
\caption{(a) Schematic of an encapulated hBN/graphene/hBN field-effect transistor on SiO$_2$/Si with a top floating gate electrode.  
(b) Diagram of the capacitive coupling between the various layers of the device structure.  
(c) Experimentally measured conductance versus gate voltage loops under gate voltage sweeps with varying periods: 86 s, 171 s, and 343 s.  
(d) Experimentally measured conductance versus gate voltage loops under different backgate sweeping from $+/-$ 10 to $+/-$ 40 V (e) Gate current measured across different periods of the applied voltage. (f) The gate capacitance extracted from the full voltage sweep. In all measuments, the top electrode is kept open, acting as a floating gate.  
}
    \label{fig:04}
\end{figure}

Interestingly, all the dynamic behaviors discussed so far break down in the presence of a FG positioned above the graphene conductive channel, as illustrated in Fig.~\ref{fig:04} (a). The FG is an electrically isolated electrode that can store charge in a non-volatile fashion and is capacitively coupled to the various layers of the transistor, as schematically depicted in Fig.~\ref{fig:04}(b). This stored charge effectively modulates the conductance state of the device, enabling it to function as a memory cell~\cite{wang2019new,Rodder2020,Wang2024,Zheng2025,Soliman2025}.

An experimental response of such a FG-graphene-based device is presented in Fig.~\ref{fig:04} (c), where the conductance loops exhibit a complete insensitivity to the gate voltage sweep period, even as it is varied over a wide range (from 86~s to 343~s). In sharp contrast, the device shows a strong sensitivity to the amplitude of the applied gate voltage, as evidenced in Fig.~\ref{fig:04}(d). This behavior is consistent with reports on similar FG device architectures based on MoS$_2$ channels~\cite{wang2019new}. Here, hBN/graphene/hBN heterostructure devices were fabricated on a SiO$_2$ (300 nm)/Si substrate, incorporating a local top FG electrode covering the source and drain regions and the electrical measurements were performed at 300 K. See the Methods section for further experimental details.

This qualitative shift in behavior is a distinctive hallmark of non-volatile memory operation and highlights the fundamentally different charge transfer dynamics induced by the FG. Remarkably, the experimental response exhibits no signatures of abrupt transitions that might be attributed to sharp tunneling events, hot-electron injection, or ferroic phase changes~\cite{park2022charge}. Despite being decoupled from the speed of the external drive, the device response consistently exhibits a dephasing offset that is notably frequency-independent. This observation poses a fundamental question: how can such behavior be reconciled within a dynamic model that accounts for charge trapping and reabsorption processes?

Initial insights are provided by the measured gate current, $J_g$, under periodic triangular voltage sweeps as displayed in Fig.~\ref{fig:04} (e) while the top electrode is floating. This current component is purely displacement in nature (note the dependance on voltage period), yet clearly reveals traces of interference from a bistable electric polarization dynamic. Further details on the expected displacement current behavior due to polarization fluctuations in layered dielectrics are provided in the Supporting Information.

Consequently, the gate capacitance, determined from the measured gate current and the applied voltage sweep rate 
\begin{equation}
C=\frac{J_g}{dV_g/dt},
\end{equation}
and presented in Fig.~\ref{fig:04} (f), displays the expected ``bow-tie'' shape (shifted up by a geometric capacitance contribution). This distinctive profile is a direct consequence of the bistable polarization dynamics, as elaborated in the Supporting Information. The capacitance, $C=C_0+C_{eff}$ therefore, is not solely dependent on the instantaneous gate bias ($V_g$) but also on the direction of its variation, thus expressed as
\begin{equation}
C_{eff} = S' \frac{d \delta P}{dV_g}=C_{eff}\left[ \text{sign}\left(\frac{dV_g}{dt} \right) V_g \right],
\end{equation}
for symmetric voltage sweeps. Additional modeling details can be found in the Supporting Information.

\begin{figure}[!htb]
    \centering
\includegraphics[width=1\linewidth]{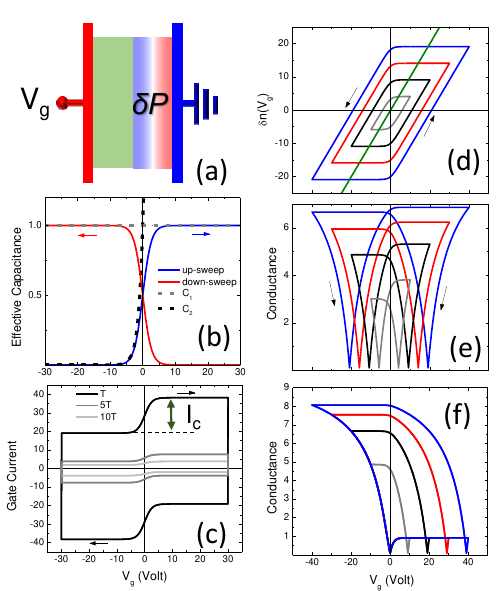}
\caption{(a) Schematic of a capacitor featuring two dielectric materials, one exhibiting polarization fluctuation. (b) Effective dynamic capacitances for forward (blue solid) and backward (red solid) voltage sweeps, alongside their constitutive stationary components (dashed curves). (c) The displacement current calculated across a range of periods for the applied gate voltage, highlighting frequency dependence. (d) The corresponding charge fluctuation versus gate voltage for different amplitudes. (e) The resultant conductance loops versus gate voltage. (f) Example of symmetric conductance loops derived from the model with fixed initialization conditions, shown for various voltage amplitudes.
}
    \label{fig:05}
\end{figure}

We can reconcile these observations with a capacitive model that incorporates an intrinsic polarization fluctuation, $\delta P$, as schematically represented in Fig.~\ref{fig:05} (a). The core assumption is that the trapping process is instantaneous, meaning no relaxation time is involved, and is primarily driven by the displacement current associated with a capacitive flux ($dV/dt$-triggering~\cite{Sze:Thyristors}). Here, the charging (dynamic doping) rate or charging current, $J_c$, is directly proportional to the rate of change of the gate voltage $\frac{dV_g}{dt}$
\begin{equation}
   J_c \equiv \frac{d \delta n}{dt} = C_{eff}\left[ \text{sign}\left(\frac{dV_g}{dt} \right) V_g \right] \frac{d V_g}{dt}.
    \label{capac}
\end{equation}
To emulate the gate voltage dependence of charging and discharging, and potentially incorporate a threshold voltage, $V_{th}$, for triggering, the effective capacitance can be simplified as two capacitors in series: $1/C_{eff}(V_g)=1/c_1+1/c_2(V_g)$. This comprises a constant geometric capacitance, $c_1$, and a depletion layer-like capacitance~\cite{Sze:Capacitors}, $c_2(V_g)=\exp{\frac{e(V_g-V_{th})}{k_B T}}$. These individual components are represented by dashed lines in Fig.~\ref{fig:05} (b), where we set $V_{th}=0$ for simplicity. The dynamic effective capacitance for both forward and backward voltage sweeps is also plotted in Fig.~\ref{fig:05} (b), showing the expected ``bow-tie'' shape, which qualitatively matches the highlighted effective capacitance step measured at the gate capacitance in Fig.~\ref{fig:04} (f) for a symmetric gate voltage sweep.

By considering the total gate current as the sum of a constant geometric contribution and the charging current in Eq.~\ref{capac}, we get
\begin{equation}
    J_g=C_0\frac{dV_g}{dt} + J_c(V_g).
\end{equation}
This model yields results that qualitatively match the gate current response previously displayed in Fig.~\ref{fig:04} (e).

Integrating Eq.~\ref{capac} yields the charge fluctuations as a function of gate voltage (Fig.~\ref{fig:05} (d)), which in turn produce the conductance loops shown in Fig.~\ref{fig:05} (e). These solutions remain entirely independent of the voltage drive speed, mirroring the experimental observations in Figs.~\ref{fig:04} (c) and (d). In this simplified dynamic response, charge carrier generation occurs through electric field variations, specifically via displacement current and capacitive injection.

It is important to note that while the solution obtained by integrating Eq.~\ref{capac} under dynamic capacitance conditions is stable and frequency independent, it relies on an initialization condition that the model itself cannot determine and must be externally imposed. Consequently, the symmetric response seen in Fig.~\ref{fig:05} (e), obtained for $\delta n (0)=c_1(V_g^{max}-V_{th})/2$ can become asymmetric by simply changing the initial condition to $\delta n(0)=0$, as demonstrated in Fig.~\ref{fig:05} (f).

We should emphasize that the agreement between the theoretical results [Figures~\ref{fig:05} (b), (c), (e)] and the experimental data [Figures~\ref{fig:04} (f), (e), (c)] goes well beyond a merely qualitative resemblance. Despite the simplicity of the model, it robustly corroborates the intrinsic correlation between gate and drain currents, as well as the crucial role of the gate current in charging the FG. Specifically, the model identifies the charging current with the step observed in the gate current [Fig. ~\ref{fig:05}~(c)]. By integrating this component from the experimental gate current [Fig.~\ref{fig:04}~(e)] over half a period, we obtain a stored charge of approximately 212 pC. Considering the effective capacitance $C_{\text{eff}} \approx 6$ pF measured in Fig.~\ref{fig:04}~(f), this charge should induce a voltage shift of about 18 V at the capacitor’s mid-point. Remarkably, this prediction, extracted solely from the experimental gate current, coincides with the observed $\Delta V_{mw}/2\approx 20$ V shift of the neutrality point in the drain current shown in Fig~\ref{fig:04}~(c).

It must also be noted that, beyond instantaneous, non-volatile responses, some systems might exhibit apparent non-volatile dynamics. In these cases, the relaxation processes are intricately linked to the frequency of the external driving force at high enough frequencies. This behavior is typical of complex systems where factors like charge trapping, interface states, diffusion processes \cite{LopezRichard2024,Warburg1901}, and non-linear dielectric responses~\cite{Lindner1962} all contribute to the system's dynamics. However, unlike true non-volatile responses, the memory effect collapses at low frequencies. More detailed information on these scenarios can be found in the Supplementary Information. 

In summary, the fundamental difference between both systems - GFETs with and without a top floating gate - lies in whether their dynamic response and charge generation are governed by inherent material properties or by the frequency of an external driving force. One system is characterized by an intrinsic relaxation time and Ohmic drift for carrier transfer, ideal for uses requiring transient responses such as pulse integration for neuromorphic applications. Conversely, the second system, with its frequency-independent charge generation via displacement current and capacitive injection, is better suited for high-frequency electronics, advanced capacitive sensing, and memory effects. Our phenomenological model was intentionally constructed in capacitive terms to highlight the experimentally observed correlations between gate displacement currents, back-gate driving conditions, and channel conductance. In this sense, the model does not explicitly exclude FN tunneling or other possible microscopic charge transfer processes such as thermionic emission, or trap-assisted tunneling, but rather incorporates them effectively through dynamic parameters (effective capacitance and relaxation times). Finally, although our experimental validation employed a graphene channel and a gold top FG, the proposed model is readily adaptable to interpret recent observations of non-volatile behavior in a wide range of two-dimensional device architectures~\cite{Liu2021,Lu2024,Sasaki2021,fu2025non}.

\section{Methods}

Both sets of GFETs were fabricated using dry transfer techniques: either by stacking a graphene monolayer onto an hBN/SiO$_2$ (300 nm)/Si substrate or by fully encapsulating the graphene between two hBN flakes~\cite{Cadore2024}. Standard plasma etching and electron-beam lithography were employed to define the floating gate (FG), source (S), and drain (D) electrodes, followed by electron-beam evaporation of Ti/Au (5 nm/50 nm). The fabrication process was completed using a conventional lift-off method.

 The first set of GFETs without a FG was measured as follows: electronic measurements were performed in a linear four-terminal configuration using standard lock-in techniques at a frequency of 17 Hz with a current bias of 100 nA. All measurements were carried out in a Janis ST-300SH cryostat system at 500 K under a pressure of $1 \times 10^{-6}$ Torr. For gate hysteresis measurements, the back-gate bias was swept in the following sequence: starting from zero volts, the voltage was swept to the maximum negative $V_g$ by a code script, then to the maximum positive $V_g$, collecting data throughout, and finally back to the negative $V_g$, again collecting data, before returning to zero volts by a code script. Due to experimental limitations, we were not able to collect the entire loop in a single continuous sweep. 

The GFETs with a FG electrode were measured as follows: transfer curves were obtained in a two-terminal configuration using standard DC techniques with a Keithley 4200A-SCS Parameter Analyzer and 100mV as constant souce-drain bias ($V_{DS}=100$ mV). All measurements were carried out in a Lake Shore Model Cryogenic Probe Station at 300 K under a pressure of $1 \times 10^{-6}$ Torr. For hysteresis loops, the back-gate bias was swept continuously from zero volts to the maximum negative $V_g$, then to the maximum positive $V_g$, back to the negative $V_g$, and finally returned to zero volts. Data acquisition (i.e. drain and gate currents as functions of the gate voltage) was performed throughout the entire voltage sweep. The top electrode positioned over the graphene channel serves as a FG, effectively acting as a charge-trapping layer~\cite{Kim2019,wang2019new,Rodrigues2025}.

\section*{Acknowledgments}
This study was financed in part by the Coordenação de Aperfeiçoamento de Pessoal de Nível Superior - Brazil (CAPES) and the Conselho Nacional de Desenvolvimento Científico e Tecnológico - Brazil (CNPq) Projs. 311536/2022-0 and 301145/2025-3, and the Sao Paulo Research Foundation (FAPESP, grant numbers 2019/14949-9 and 2023/09395-0), and through the Research, Innovation and Dissemination Center for Molecular Engineering for Advanced Materials – CEMol (Grant CEPID No. 2024/00989-7). The authors also acknowledge the Brazilian Nanotechnology National Laboratory (LNNano) and Brazilian Synchrotron Light Laboratory (LNLS), both part of the Brazilian Centre for Research in Energy and Materials (CNPEM), a private nonprofit organization under the supervision of the Brazilian Ministry for Science, Technology, and Innovations (MCTI), for device fabrication and electrical characterization - LNNano/CNPEM (proposals: MNF-20250043, MNF-20250045, MNF-20240039, and MNF- 20240040) and LAM-2D (proposals: 20240497) at LNLS/CNPEM, besides Marcelo R. Piton for the experimental assistance. K.W. and T.T. acknowledge support from the JSPS KAKENHI (grant numbers 21H05233 and 23H02052) and World Premier International Research Center Initiative (WPI), MEXT, Japan.

\bibliography{arxiv.bbl}
\end{document}